\documentclass[a4paper]{article}

\usepackage{INTERSPEECH2022}
\usepackage{subfigure}

\title{Boosting Tail Neural Network for Realtime Custom Keyword Spotting}
\name{Sihao Xue$^1$, Qianyao Shen$^1$, Guoqing Li$^1$}
\address{
  $^1$Metawall}
\email{sihao.xue@metawall.ai, qianyao@kth.se, guoqing.li@metawall.ai}

\begin{document}

\maketitle
\begin{abstract}
    In this paper, we propose a Boosting Tail Neural Network (BTNN) for improving the performance of Realtime Custom Keyword Spotting (RCKS) that is still an industrial challenge for demanding powerful classification ability with limited computation resources. Inspired by Brain Science that a brain is only partly activated for a nerve simulation and numerous machine learning algorithms are developed to use a batch of weak classifiers to resolve arduous problems, which are often proved to be effective. We show that this method is helpful to the RCKS problem. The proposed approach achieve better performances in terms of wakeup rate and false alarm. 
    
    In our experiments compared with those traditional algorithms that use only one strong classifier, it gets 18\% relative improvement. We also point out that this approach may be promising in future ASR exploration. 


  

\end{abstract}
\noindent\textbf{Index Terms}: Boost Tail Neural Network, Realtime Custom Keyword Spotting, On Device

\section{Introduction}

In recent years, automatic speech recognition (ASR) is widely applied in various thriving fields such as automobiles. To avoid massive resource consumption and user intent misunderstanding caused by ASR and natural language understanding, keyword spotting (KWS) is usually used to trigger followed ASR task. Therefore, it is required that the KWS system has special properties such as high wakeup rate, low false alarm rate, and low latency stream compute pattern. In the case of automobile, spotting task generally runs on a low compute resource device, so the development of a spotting algorithm that consumes low computational cost without sacrificing the performance is strongly motivated.

Traditionally, keywords are given before the project launch, which largely simplifies the problem because developers are able to do targeted optimization in various ways for the fixed keywords. For this type of problem, much work has been proved effective [1][2]. However, the demand of more flexibility is increasing with the growth of complication of application situations. Therefore, RCKS, is now more preferable because it allows the users to define the keywords by themselves. As it is next to impossible to foresee users' ideas before the project launch, RCKS model must have classification ability for all acoustic states to identify the keywords spoken by the users. Unlike the situations of ASR where logical and grammatical sentences are often used, RCKS obtain little assistance from language models because only a word or a short phrase is used in most cases, which implies that acoustic models play an essential role in such a case. The existing algorithms for RCKS, however, cannot fully satisfy the requirement of high acoustic classification accuracy with low computation cost and latency. 
The main purpose of this paper is to design a BTNN to improve the performance of RCKS to achieve the better classification accuracy with much lower computational consumption and latency. 

\section{Boost Tail Neural Network}


\subsection{Feature Embedding}
In neural networks , the output from on hidden layer can be regarded as the input features for the next. It is similar to [3][4]. In the case of our interest in this paper, more specifically, the abstract features in the output of some hidden layer can be treated as a usually nonlinear transformation of the original acoustic feature, the serve as the input for the subsequent layers to perform classification. Theoretically, the output of deeper hidden layers exhibits classification, which, therefore, we call feature embedding. One extreme example is that only a linear transformation is needed to apply on the embedding results from the last hidden layer before performing the softmax classification. Figure 1 shows the features embedding for some acoustic states. there is obvious classification, although the physical meaning is difficult to explain analytically. 

\begin{figure}
\centering
\subfigure{
\begin{minipage}[b]{0.2\textwidth}
\includegraphics[width=1\textwidth]{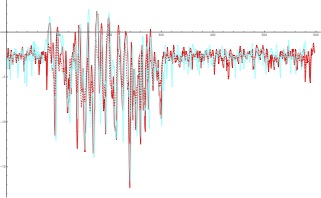} 
\end{minipage}
}
\subfigure{
\begin{minipage}[b]{0.2\textwidth}
\includegraphics[width=1\textwidth]{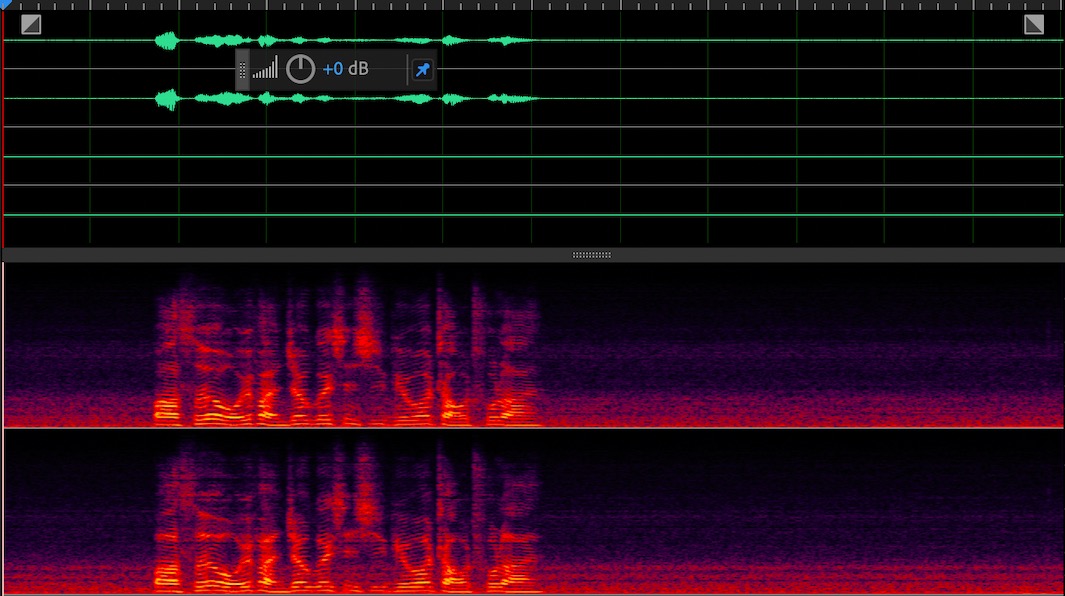} \\
\end{minipage}
}
\caption{Feature Embedding and original audio info. The pattern of embedding matches with audio info in time domain, although the pattern is difficult to explain analytically. }

\end{figure}




\subsection{Boosting Tail Neural Network}
In machine learning, one common method is the boosting learning [5][6][7] which uses batch of weak classifiers in stead of one strong classifier to perform mode matching calculation, while each of the weak classifiers is only capable of simple classification. Combining with the feature embedding mentioned above, we propose to use multiple weak classifiers on the features embedding to improve the performance of classification, which we call BTNN ( borrow the name from boosting algorithm, shown in Figure 2). More precisely, BTNN consists of a shared feature embedding extractor and numerous binary classifier, each of which is used to calculate a specific modeling unit. Subsequently, we can obtain the probability of each acoustic state for the corresponding binary classifier. Since each classifier is trained independently,  we can apply MSE loss to the training process of each acoustic state. 

\begin{equation}
  loss_{i} = 
  \begin{cases}
  MSE(o_{s_{i}}, 0) & \text{for negative samples} \\
  MSE(o_{s_{i}}, 1)\times S_{i} & \text{for positive samples}
  \end{cases}
  \label{eq1}
\end{equation}

To maintain balance of positive and negative samples of each state, different scale coefficient \(S_{i}\) is applied for positive and negative samples as in (\ref{eq1}). 

\begin{figure}[t]
  \centering
  \includegraphics[width=\linewidth]{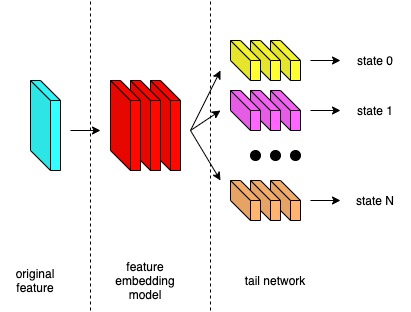}
  \caption{BTNN. The feature embedding model can be any type model. The tail model should be feedforward model without time dependence}
  \label{fig:fe}
\end{figure}

For BTNN, each classifier focuses on one particular acoustic state. Hence, in the training process, each state can have targeted training parameters to optimize its classification ability. For example, due to the different amount of data of different acoustic states in the training data, we can use different ratios of positive and negative samples for training to improve the resulting performance of the classifiers.

\section{RCKS post-processing}
\subsection{acoustic post-processing}

The output of the model is not probability because of the MSE loss in (\ref{eq1}). Therefore it is not suitable to push the model output to the decode part directly. To resolve this problem, we need to analyze the statistics and probability distribution of some development data with alignment (the environment fit data that covers all acoustic states will be better) as shown in Figure 3. This result will be used for the computation of static probability in the whole RCKS process.

\begin{figure}[t]
  \centering
  \includegraphics[width=8cm,height=6cm]{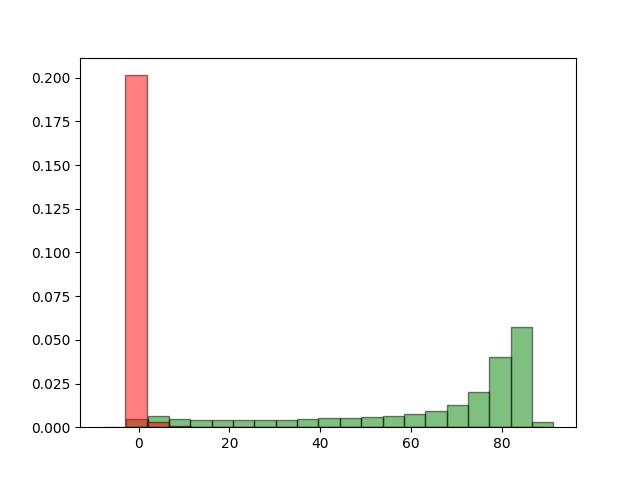}
  \caption{State distribution. The overlap area may represent homologous pronunciation, such as "z" and "zh" in Chinese}
  \label{fig:fe}
\end{figure}

Different from the Softmax output, the sum of the resulting "probability" from BTNN may not be 1 because there can be coupling among different states, as shown in Figure 4. One of the advantages of this methods is that it takes into account the relation of similar acoustic states, which helps to reduce the acoustic confusion resulting from accents or originally similar pronunciation. 

First, according to Figure3, the distribution can be analyzed with a batch of data by segmentation and interpolation. The range of each state output can be split in several segmentation and the probability of each boundary can be computed (\ref{eq2}) shown in Figure 5. 

\begin{equation}
  P_{n} = 
  \begin{cases}
  0 & \text{for n = 0} \\
  1 & \text{for n = N} \\
  P_{n+1} - \frac{C_{n+1}}{C_{total}} & \text{for 1 \(\leq\) n \(\leq\) N - 1}
  \end{cases}
  \label{eq2}
\end{equation}

In forward process, two probability (positive probability and negative probability) can be computed (\ref{eq3}, \ref{eq4}). For the computation of finally probability, we need to use the geometric distribution with scaling coefficients which are required to balance positive and negative scores (\ref{eq5}).The parameter \(S_{s,p}\) and \(S_{s,n}\) depends on discrimination of each state, these two parameter can be set manual or computed by some algorithms. In our experiment, besides setting \(S_{s,p} = 4\) \(S_{s,n} = 1\), we compute \(S_{s,p}\) \(S_{s,n}\) to maximum frame level likelihood for each state based on some dev-data.Since the classifications for different states are independent in the model, we have freedom to choose the optimize way for each state. 

\begin{equation}
  p_{p}(x) = 
  \begin{cases}
  0 & \text{ x \(\leq\) \(b_{0}\)} \\
  1 & \text{ x \(\geq\) \(b_{N}\)} \\
  P_{n} + \frac{(x-b_{n}) \times (P_{n+1}-P_{n})}{b_{n+1} - b_{n}} & \text{ \(b_{n}\) \(\leq\) x \(\leq\) \(b_{n+1}\)}
  \end{cases}
  \label{eq3}
\end{equation}

\begin{equation}
  p_{p}(x) = 
  \begin{cases}
  1 & \text{ x \(\leq\) \(b_{0}\)} \\
  0 & \text{ x \(\geq\) \(b_{N}\)} \\
  P_{n} - \frac{(x-b_{n}) \times (P_{n}-P_{n+1})}{b_{n+1} - b_{n}} & \text{ \(b_{n}\) \(\leq\) x \(\leq\) \(b_{n+1}\)}
  \end{cases}
  \label{eq4}
\end{equation}

\begin{equation}
  p_{s}(x) = (p_{s,p}(x)^{S_{s,p}} \times p_{s,n}(x)^{S_{s,n}})^{\frac{1}{S_{s,p}+S_{s,n}}}
  \label{eq5}
\end{equation}

\begin{figure}[t]
  \centering
  \includegraphics[width=4cm,height=3cm]{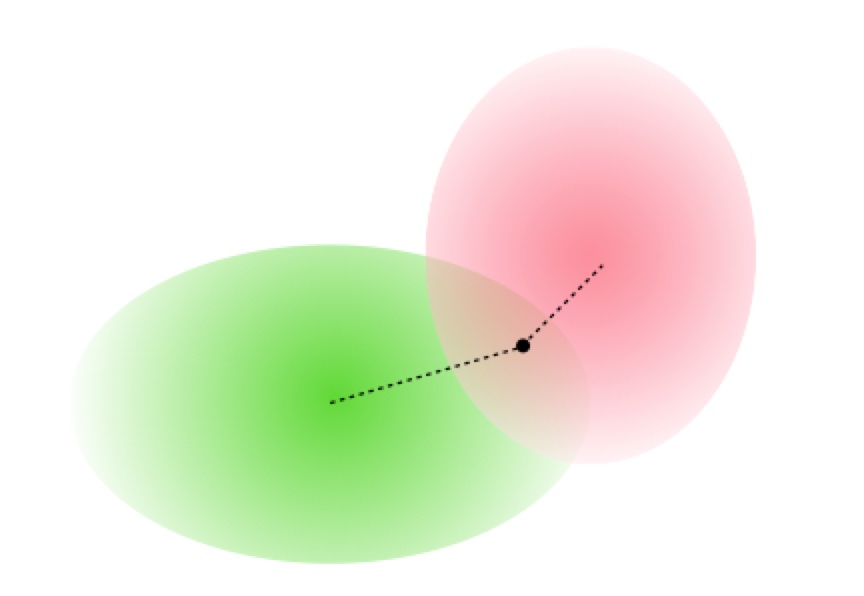}
  \caption{The probability is computed with each distribution independently, therefore there is no \(\sum p_{i} = 1\) constraint. There may be several high probability state for acoustic similarity or no high probability for pronunciation that out of modeling }
  \label{fig:fe}
\end{figure}

\begin{figure}[t]
  \centering
  \includegraphics[width=\linewidth]{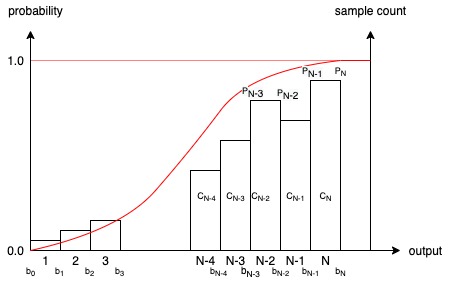}
  \caption{For segmentation \(N\), the number of samples is \(C_{n}\), the probability of each boundary is \(P_{n}\)}
  \label{fig:fe}
\end{figure}

\subsection{decode processing}

Token push [8][9] is selected to do decode part. When user set his own keyword, the system will build a Weighted Finite-State Transducer (WFST) [10] according to the lexicon. In the WFST, not only the traditional arc but jump arc can be used to resolve the acoustic jump situation where a few wrong acoustic score computation leads to the failure of the whole recognition flow. Nevertheless, jump arc punishment score is necessary to reduce the false alarm rate which is shown is Figure 6.

Moreover, the required acoustic sate is easy to get the active token in the decoding processing. In the next frame of computation, only particular acoustic score needs to compute. The compute flow is shown in Figure 7. At \(t_{n}\), get required states, self loop state (\(state_{slp}\)) and emission state (\(state_{em}\)), from active token and activate corresponding tail net. This computation way cleans up uncorrelated acoustic state score computation. So pure feedforward architecture is chosen for tail part. For Feature Embedding part, some time-rely architecture can be chosen to improve model ability.

\begin{figure}[t]
  \centering
  \includegraphics[width=\linewidth]{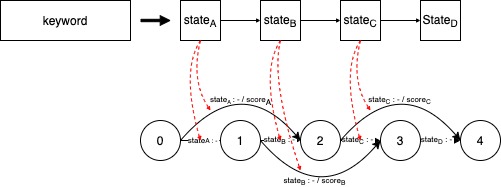}
  \caption{WFST building pattern. Be different from normal arc, jump arc has punishment score to restrain false alarm.}
  \label{fig:fe}
\end{figure}

\begin{figure}[t]
  \centering
  \includegraphics[width=\linewidth]{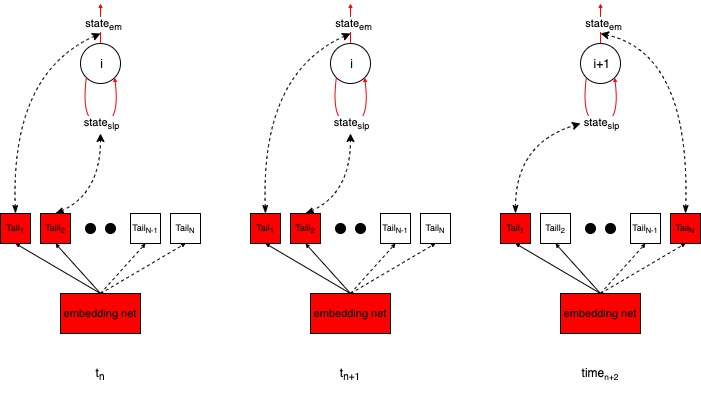}
  \caption{Recognition flow. The red part actually participates in computation flow at each frame}
  \label{fig:fe}
\end{figure}

\section{Experiment}

We use a datasets combined with 3000 hours vehicle data  and we do not exclusively add any keywords data into these datasets to ensure that the experiment is about RCKS process. Besides splitting datasets into training and validation for training process, we select 10k utterances randomly for computing \(S_{s,p}\) \(S_{s,n}\)  to maximum frame level likelihood for each state . The test data includes positive (5000 utterances with 40 keywords include Chinese vehicle commands word (include "FANG DA DI TU", "ZENG DA YIN LIANG" and etc) and wakeup words (include "XIAO WEI TONG XUE", "LAO WEI TONG ZHI" and etccd) defined by users) and negative (50 hours normal speech without any keywords) data. All test data is recorded in automobiles.  Data augmentation was performed using automobiles simulator includes adding car-type noise and Acoustic Echo Cancellation (AEC) simulation. The noise sources were also recorded in automobiles.

The base model is a 8-layer DFSMN ( 128 hidden layer and 512 memory block dim) [11] and two projection layers (128 DNN layer) which has about 1.2M computation and both of its hardware consumption and computation speed is acceptable on 8155 vehicle chip. Further more, we also trained a 10-layer DFSMN with 1.5M computation for comparison. For BTNN, embedding net is same as base model's FSMN part, each tail net architecture is pyramid architecture \(128 \times 64 \times 32 \times 1\). Asynchronous Stochastic Gradient Descent (ASGD) and Adam optimizer was used for training our model. And we choose 40-dim Filterbank to be input acoustic feature. 

\subsection{Constant Scale BTNN}
As described in 3.1, we first set \(S_{s,p} = 4.0\) and \(S_{s,n} = 1.0\) for computing acoustic post-processing score. The result is shown in Table1. 

\begin{table}[th]
  \caption{For several parameters can be adjusted, the tables only shows the best wakeup rate with 1 false alarm per 24 hours}
  \label{tab:table1}
  \centering
  \begin{tabular}{ r@{}l  r }
    \toprule
    \multicolumn{2}{c}{\textbf{Model}} & 
                                         \multicolumn{1}{c}{\textbf{Wakeup Rate}} \\
    \midrule
    $8-layer FSMN$                       & $ $ & $89.14\%$~~~             \\
    $10-layer FSMN$                       & $$  & $90.40\%$~~~               \\
    $csBTNN$                       & $$  & $ 91.10\%$~~~       \\
    \bottomrule
  \end{tabular}
\end{table}

At the same false alarm level, csBTNN achieves relative 18\% improvement compared-to 8-layer FSMN. Besides this, csBTNN also represents better than 10-layer FSMN which has large computation consumption. 

\subsection{Adaption Scale BTNN}
Besides csBTNN, we also try to apply different \(S_{s,p}\) and \(S_{s,n}\) for each state. As describe in 3.1, we first compute \(p_{p}(x)\) and \(p_{n}(x)\) and compute the likelihood (\(like_{s}(x)\)) between alignment with final acoustic scores according to several \(S_{s,p}\) and \(S_{s,n}\). Because of the unbalanced data proportion for each state, we set different scaling for positive (0.99) and negative (0.01) samples (\ref{eq6}). The result is shown in Table 2.

\begin{equation}
  like_{s}(x)=
  \begin{cases}
  0.99 * like(ali_{x}, p_{s}(x)) & \text{ x is pos sample} \\
  0.01 * like(ali_{x}, p_{s}(x)) & \text{ x is neg sample} 
  \end{cases}
  \label{eq6}
\end{equation}

\begin{table}[th]
  \caption{}
  \label{tab:table2}
  \centering
  \begin{tabular}{ r@{}l  r }
    \toprule
    \multicolumn{2}{c}{\textbf{Model}} & 
                                         \multicolumn{1}{c}{\textbf{Wakeup Rate}} \\
    \midrule
    $8-layer FSMN$                       & $ $ & $89.14\%$~~~             \\
    $10-layer FSMN$                       & $$  & $90.40\%$~~~               \\
    $csBTNN$                       & $$  & $91.10\%$~~~       \\
    $asBTNN$                       & $$  & $91.64\%$~~~       \\
    \bottomrule
  \end{tabular}
\end{table}

Compared with csBTNN, asBTNN achieve about 6\% relative improvement. 

\section{Prospect}
In this paper, we propose to use BTNN to save computation consumption. The model size is actually large, e.g in our experiment the model size is larger than 2M, but the frame-level computation can be significantly reduced to a similar level of a small one by controlling activation of acoustic state. This algorithm is inspired by the work pattern of our brains, i.e. different parts responsible for different problems. The embedding part of the network can then be regarded as fore-end organs such as those transform the acoustic signals into electric signals in the nerves.

Be different from softmax model, as shown in Figure 8, the output of tail network allows state space to be overlapping with each other, therefore it is easier to emerge similarity between states and deal with different manner of articulation such as the stretched voice. On the contrary, the posterior result of softmax model is usually instable at this situation. 

Meanwhile, this way of processing profits acoustic analysis. When people pronounce a word in different contexts, the acoustic states also change with separate pronunciation. This changing processing can be seen as combination and change of multi normal modes, which is more physically intuitive. For those states that strongly couple with each other or have low distinction, we can train the system for them independently. 

\begin{figure}[t]
  \centering
  \includegraphics[width=\linewidth]{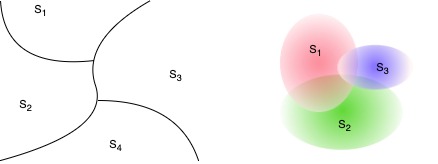}
  \caption{Difference between non-overlapping and overlapping model}
  \label{fig:fe}
\end{figure}

\section{Conclusions}

We have presented Boosted Tail Neural Network (BTNN) and post-processing algorithms to resolve Realtime Custom Keyword Spotting (RCKS). The embedding neural network can be any network architecture although only FSMN is used in this paper. Therefore, to future improve result, we will train more architecture e.g [12][13][14] to better performance. Besides this, the overlapping property may be able to used for acoustic pattern analysis more physically and we will try to find relation through vibration normal modes and acoustic states.   



\bibliographystyle{IEEEtran}

\bibliography{mybib}

\end{document}